%% file: main.tex
\documentclass[lettersize,journal]{IEEEtran}
\usepackage{amsmath,amssymb,amsfonts}
\usepackage{algorithmic}
\usepackage{algorithm}
\usepackage{array}
\usepackage[caption=false,font=normalsize,labelfont=sf,textfont=sf]{subfig}
\usepackage{textcomp}
\usepackage{stfloats}
\usepackage{url}
\usepackage{verbatim}
\usepackage{graphicx}

\usepackage{cite}

\usepackage{comment}
\usepackage[dvipsnames]{xcolor}
\usepackage{enumerate}
\usepackage{enumitem}
\usepackage{makecell,tabularx, float}
\usepackage{soul}
\usepackage{multirow}
\usepackage[pagebackref=true]{hyperref}
 \hypersetup{ %
  colorlinks=true, %
  pdfstartview={FitH}, %
  citecolor=Black, %
  linkcolor=Black, %
  urlcolor=Blue %
 }

\begin{document}

\title{A Distributed Approach for Agile Supply Chain Decision-Making Based on Network Attributes}

\author{Mingjie Bi,~\IEEEmembership{Student Member,~IEEE}, Dawn M. Tilbury,~\IEEEmembership{Fellow,~IEEE}, Siqian Shen,~\IEEEmembership{Member,~IEEE}, and Kira Barton,~\IEEEmembership{Senior Member,~IEEE}
\thanks{This work was funded in part by the United States National Science Foundation (NSF) grant \#CMMI-2034974.}
\thanks{Mingjie Bi is with the Robotics Department, 
        University of Michigan, Ann Arbor, MI 48109, USA
        {\tt\small mingjieb@umich.edu}}
\thanks{Dawn M. Tilbury and Kira Barton are with the Robotics Department and the Department of Mechanical Engineering, 
        University of Michigan, Ann Arbor, MI 48109, USA
        {\tt\small \{tilbury, bartonkl\}@umich.edu}}
\thanks{Siqian Shen is with the Department of Industrial and Operations Engineering, 
         University of Michigan, Ann Arbor, MI 48109, USA
        {\tt\small siqian@umich.edu}}
}

\markboth{IEEE TRANSACTIONS ON AUTOMATION SCIENCE AND ENGINEERING}%
{Bi \MakeLowercase{\textit{et al.}}: A Distributed Approach for Agile Supply Chain Decision Making Based on Network Attributes}


\maketitle

\input{sections/0_0_abstract}

\input{sections/1_0_introduction}

\input{sections/2_0_supplychain}

\input{sections/3_0_approaches}

\input{sections/4_0_setup}

\input{sections/5_0_casestudy}

\input{sections/6_0_conclusion}

\input{main.bbl}

\input{bio/biotext}

\vfill

\end{document}

%% file: sections/0_0_abstract.tex
\begin{abstract}
In recent years, the frequent occurrence of disruptions has had a negative impact on global supply chains.
To stay competitive, enterprises strive to remain agile through the implementation of efficient and effective decision-making strategies in reaction to disruptions. 
A significant effort has been made to develop these agile disruption mitigation approaches, leveraging both centralized and distributed decision-making strategies. 
Though trade-offs of centralized and distributed approaches have been analyzed in existing studies, no related work has been found on understanding supply chain performance based on the network \textit{attributes} of the disrupted supply chain entities. 
In this paper, we characterize supply chains from a capability and network topological perspective and investigate the use of a distributed decision-making approach based on classical multi-agent frameworks. 
The performance of the distributed framework is evaluated through a comprehensive case study that investigates the performance of the supply chain as a function of the network structure and agent attributes within the network in the presence of a disruption. 
Comparison to a centralized decision-making approach highlights trade-offs between performance, computation time, and network communication based on the decision-making strategy and network architecture. 
Practitioners can use the outcomes of our studies to design response strategies based on agent capabilities, network attributes, and desired supply chain performance.
\end{abstract}

\def\abstractname{Note to Practitioners}
\begin{abstract}
This research is motivated by the challenges in determining agile decision-making strategies that enable a supply chain enterprise to adapt to disruptions while taking into account the network-based attributes of the disrupted agent and the requirements of the supply chain system. 
Existing approaches in the literature focus on providing one feasible decision-making strategy based on specific performance metrics. 
This paper investigates both centralized and distributed approaches to better understand the differences between the response strategies in the case of supplier loss. More specifically, we design a supply chain instance and conduct a case study to evaluate the performance of the centralized and distributed approaches in terms of several common performance metrics used in practice. 
The case study provides insights for users to select a decision-making approach based on the network attributes and agent capabilities of the supply chain. 
The impact of network uncertainties and risk assessment are not considered in this work. 
Future studies will investigate a stochastic supply chain environment and heterogeneous risk management framework in the context of agile decision-making for disrupted supply chain enterprises.
\end{abstract}


%% file: sections/1_0_introduction.tex
\section{Introduction}
\label{sec:intro}

\IEEEPARstart{A}{s} the complexity and uncertainty of global supply chains increase, various disruptive events (e.g., COVID-19, war, and trade regulation) have occurred more frequently and with greater intensity, severally impacting global supply chains through changes in customer demand, material supply and manufacturing capabilities~\cite{rahman2021agent, xu2020disruption}. 
Existing literature in this domain focuses on proactive methods for supply chain disruption mitigation, such as demand forecasting, inventory management, and stochastic optimization methods that estimate potential disruptions in advance to enhance supply chain robustness~\cite{ho2015supply,ivanov2017literature}. 
However, unexpected disruptive events, such as supplier loss, require enterprises to make quick and effective decisions in response to the disruption, by understanding how the selection of a mitigation approach will impact the recovery performance~\cite{ben2019internet}. 
In this paper, we consider the following problem formulation: given a supply chain network (SCN) with a specific structure, individual enterprise agents with varying attributes as a function of the network, existing product flows, and an unexpected disruption to one of the agents within the network, how do the agent attributes and decision-making strategy impact the system performance?

In the literature, researchers have described a supply chain as a network with vertices (e.g., supplier, customer, etc.) and edges (e.g., transportation), along with their associated attributes and parameters (e.g., cost and capacity)~\cite{parhi2018topological, kim2015supply}. 
In this paper, both the vertices and edges we consider have intelligence and thus are defined as agents. 
Therefore, understanding the agents and their attributes within an SCN can help determine the impact of disruptions based on where disruptions occur and the critical performance metrics of interest. 
From the topological perspective, existing literature has made a significant effort in conceptualizing supply chain disruptions and investigating the effects of the overall network topology on supply chain resilience and robustness~\cite{kim2011structural, kim2015supply, nair2011supply, zhao2022supply, estrada2023multi}. 
However, the existing studies did not discuss how the agent attributes within this network (e.g. capabilities, connectivity, etc) impact the mitigation performance. 
Furthermore, agent capability attributes are important for understanding the impact of the disrupted agent on supply chain disruption recovery. 
This specification has implications for how we determine the decision-making approach for disruption mitigation.

In terms of decision-making approaches for supply chain management, centralized models are widely used to provide optimal solutions based on specific objectives (e.g., product flow cost)~\cite{croom2000supply}. 
For disruption mitigation, centralized models require information about the entire supply chain in order to re-optimize the system in response to a disruption~\cite{fischer2020design}. 
As the complexity and scale of supply chains increase, it becomes more difficult to remain agile in the presence of multiple disruptions~\cite{baryannis2019supply} due to communication demands and computational complexities that arise in these scenarios~\cite{giannakis2016multi}. 
To improve the agility and effectiveness of SCNs, researchers have proposed distributed decision-making approaches, where multiple entities in the system make decisions via local communication and collaboration~\cite{xu2021will, toorajipour2021artificial,cavone2020design,bi2021dynamic}.

Multi-agent control is a distributed approach that can enable intelligent decision-making for agile supply chain disruption mitigation~\cite{whitbrook2017reliable,bi2023distributed}. 
Each autonomous agent in an SCN either represents a physical entity, such as a supplier, or is responsible for a function, e.g., demand forecasting~\cite{giannakis2016multi}. %
Various agents communicate and make decisions collaboratively based on their knowledge and goals~\cite{kovalenko2022toward,bi2021dynamic}. 
In the existing literature, most agent-based disruption reaction strategies are based on either pre-defined disruption scenarios and reactive actions~\cite{torabi2015resilient,blos2018framework,giannakis2011multi} or rule-based reasoning and logic~\cite{ghadimi2019intelligent, pal2014multi, wang2013ontology, shukla2016fuzzy}. 
The disruption reaction performance of these methods is limited to a set of pre-defined scenarios and rules. 
It can be difficult to include a large number of predefined scenarios or define all of the necessary rules for large and complex supply chains. 
Therefore, applying these approaches to investigate the correlations between agent attributes and system performance requires significant design efforts.
In our prior work~\cite{bi2022model, bi2023distributed}, we provide a multi-agent framework that is flexible to design and enables a dynamic disruption response in the supply chain.
The agents are equipped with a model-based architecture instead of rule-based logic.
However, it is limited to certain disruption scenarios and small-scale SCNs, and does not analyze how the network attributes of the disrupted agents affect performance. 
In summary, to the best of our knowledge, no study has carried out an evaluation of the performance of both centralized and distributed approaches based on the attributes of the disrupted supply chain agents in a complex SCN.



To address this limitation, the contributions of this paper over the previous work include:  
(1) Formulation of an agent's attributes (e.g. capability, connectivity) within the context of an SCN. 
(2) Extension of our preliminary multi-agent framework to handle different disrupted agents in complex networks by allowing agent exploration and iterative communication. 
(3) Validation of the proposed approach through an investigation of the correlations between agent attributes and performance of disruption responses with a comprehensive case study.

The rest of the paper is organized as follows. 
In Section~\ref{sec:supplychain}, we describe an SCN from topological and capability perspectives. 
In Section~\ref{sec:approaches}, we present the proposed distributed multi-agent framework for disruption mitigation. 
In Section~\ref{sec:setup}, we design the case study set-up, including a extensive supply chain instance, a centralized approach as a benchmark, and performance metrics. 
In Section~\ref{sec:casestudy}, we conduct a performance analysis of centralized and distributed approaches, with concluding remarks given in Section~\ref{sec:conclusion}.






%% file: sections/2_0_supplychain.tex
\section{Supply chain network}
\label{sec:supplychain}

\input{sections/2_1_scdisruptions}

\input{tables/notation}

\input{sections/2_2_attributes}

These attributes describe the current status of agents in terms of topology and capability and they are independent of each other.
A change of one attribute will not affect other attributes.
But these attributes may change together as the agent status changes.
In this paper, we focus on the disruption (loss) of a single agent. 
While any agent loss is important, we hypothesize that the impact of the agent loss is affected by the attributes of the specific agent. 
Note that though the attributes are dynamically updated if needed as the SCN operates, we focus on the attributes when the disruption occurs and
investigate how the attributes of a disrupted agent affect the performance of the mitigation decisions.

%% file: sections/2_1_scdisruptions.tex
\subsection{Overview and assumptions}
\label{sec:assumptions}

Consider an SCN $G(V,E)$ with $V$ being the set of vertices, representing supply chain entities, such as suppliers, customers, etc., and $E$ being the set of edges, representing product/material flows between the entities. 
We also consider the associated information (e.g., demand, production, cost, and capacity) with the vertices and edges~\cite{bi2022model,kim2015supply}. 
In this paper, both vertices and edges in the SCN have intelligence and thus are defined as agents.
The corresponding supply chain agent network $G^a(A,L)$ includes all the vertices in $V$ and edges in $E$ in the agent set $A$, and the agents in $A$ are connected by a set $L$ of communication links. 
Similar to the set-up in~\cite{bi2022model}, in this paper, we consider the following agent types in the network: customer, distributor, original equipment manufacturer (OEM), tier supplier, and transporter.

This paper investigates how the network attributes of a given agent (e.g. connectivity, capability) impact the performance of the supply chain system in response to a disruption. 
Supply chain disruptions are classified into different categories, including internal and external disruptions based on their causes, as well as supplier and customer disruptions based on their locations~\cite{ho2015supply,ivanov2017literature}. 
In this study, our primary focus is on an unexpected supplier loss, which may be triggered by natural disasters or workforce strikes.

The response is defined as a new flow plan to minimize total cost and demand dissatisfaction.
We explore both distributed and centralized decision-making approaches to provide a comparison. 
The centralized approach solves the problem from the entire SCN perspective:

\begin{subequations}
\small
\label{mip}
\begin{align}
\min_{y, x, I, p, \beta, \zeta, \Delta}& \hspace{1ex}\mathcal{J}=\sum_{(i,j) \in E, k \in K} c_{ijk}y_{ijk} +\sum_{i \in V, k \in K} e_{ik}p_{ik} \nonumber \\
& \quad + \sum_{i \in V, k \in K}\rho_{ik}^d \Delta_{ik}^d  \label{obj:mip}\\
\text{s.t.}\qquad
&\label{cst:flowbalance}
\sum_{j: (i, j)\in E}y_{ijk} - \sum_{j: (j, i)\in E}y_{jik} +  \sum_{k' \in K}r_{kk'}p_{ik'}  \nonumber\\
& \hspace{0ex}- p_{ik} = x_{ik} + I_{ik}^0 - I_{ik}, \ \forall i \in V,\ k \in K \\
\label{cst:cap_y}
&\sum_{k \in K}y_{ijk} \leq q_{ij}\beta_{ij}, \ \forall (i,j) \in E\\
  \label{cst:cap_p}
& \sum_{k \in K}p_{ik} \leq \bar{p}_{i}\zeta_{i},\ \forall i \in V\\
\label{cst:pnlty_d}
& \Delta_{ik}^d \geq x_{ik}- d_{ik},\ \forall i \in V,\ k \in K\\
&  y_{ijk}, x_{ik}, I_{ik}, \Delta_{ik}^d \geq 0,\ \zeta_{i}, \beta_{ij} \in \{0, 1\},\ \nonumber\\
\label{cst:domain}
& \hspace{13ex}\forall i \in V,\ (i, j)\in E,\ k \in K,
\end{align}
\end{subequations}
where \eqref{obj:mip} are the total costs of flow, inventory, and production, as well as penalty costs of unsatisfied demand.
Constraint in~\eqref{cst:flowbalance} defines flow balance of each product for each agent; constraint in~\eqref{cst:cap_y} and~\eqref{cst:cap_p} limits the flow on each edge and production at each agent by its given capacity; constraint in~\eqref{cst:pnlty_d} computes the unsatisfied demands of each product at each vertex; and constraint in~\eqref{cst:domain} specifies the domains of variables.
Once the disruption is identified, a centralized decision-maker will re-run the centralized model with updated network structures, parameters, and constraints to determine the re-optimized decisions.
More details can be found in our previous work~\cite{bi2022model}.

To describe the proposed distributed approach, we firstly make the following assumptions to specify our scope:
\begin{enumerate}[label={A.\arabic*}]
    \item Supply chain agents have self-awareness of their own attributes and can communicate and make decisions.\label{a1}
    \item The SCN or system contains supplier redundancy and operates within the capacity limit of the initial plan. \label{a2}
    \item Unexpected disruptions are in the form of a lost agent (e.g. agent is unable to perform their set tasks) within the SCN. This disruption can be detected by the associated agent. \label{a3}
    \item An agent retains communication capabilities even in the presence of a disruption.  \label{a4}
\end{enumerate}
\ref{a1} defines the agents' abilities to make individual or locally dependent decisions in response to a disruption. 
\ref{a2} ensures that a new product flow plan can be determined. 
\ref{a3} guarantees that the disruption will be identified by the agent if it occurs and also designates how the SCN will be impacted by the disruption. 
\ref{a4} is necessary to enable local negotiations in response to the disruption. 

To understand how disruptions affect supply chains, we provide detailed supply chain descriptions at the network and agent levels. 
Specifically, we focus on the role of each agent in the SCN from both topological and capability perspectives. 
Table~\ref{tab:notation} summarizes the notations used in this paper.
We also illustrate key definitions in Figure~\ref{fig:egnetwork}.





%% file: tables/notation.tex
\begin{table}[tb]
\caption{Nomenclature}
\label{tab:notation}
    \begin{tabularx}{\columnwidth}{lX}
    \hline
    \multicolumn{2}{l}{\textbf{Supply chain description}}\\
    \hline
    $G=(V, E)$ & supply chain network (vertices and transportation edges).\\
    $G^a=(A, L)$ & agent network (agents and communication links).\\
    $K$ & set of product types.\\
    $y_{ijk}$ & units of product $k$ transported from agent $a_i$ to $a_j$.\\
    $f$ & network flow state including all product flows.\\
    $m=(o,k)$ & capability of performing operation $o$ for product $k$.\\
    $Z(k)$ & the set of needed product types to make a product $k$.\\
    \hline
    \multicolumn{2}{l}{\textbf{Agent attributes}}\\
    \hline
    $C_i$ & connectivity of agent $a_i$.\\
    $D_i$ & depth of agent $a_i$ in the network.\\
    $R_i(m)$ & capability redundancy of agent $a_i$ for capability $m$.\\
    $P_i$ & production complexity of agent $a_i$.\\
    \hline
    \multicolumn{2}{l}{\textbf{Agent communication and decision-making}}\\
    \hline
    $a_e$ & disrupted (i.e., lost) agent\\
    $y_e^0$ & all the initial product flows related to $a_e$.\\
    $A_{dm}$ & set of demand agents\\
    $d_{ik}$ & units of product $k$ that agent $a_i$ needs.\\
    $\Delta f$ & changes of network flow state.\\
    $\mathcal{M}_j(k)$ & set of agents that $a_j$ sends requests to for product $k$.\\
    $\Bar{y}_{ijk}$ & maximum units of product $k$ that agent $a_i$ determines to provide to $a_j$.\\
    $y'_{ijk}$ & units of product $k$ that agent $a_j$ determines to get from $a_i$.
    \\
    \hline
    \multicolumn{2}{l}{\textbf{Metrics for performance evaluation}}\\
    \hline
    $O$ & sum of the costs for transportation and production that exceed the nominal agent capacity.\\
    $N_c$ & sum of modified edges (e.g. type and/or amount of production flow) and agents (e.g. new production volume or capability).\\
    $N_a$ & sum of additional edges and agents needed for transportation and production.\\
    $M$ & the number of agent communication exchanges used to derive a response to disruption.
    \vspace{0.3em}
    \\
    \hline
    \end{tabularx}
\end{table}

%% file: sections/2_2_attributes.tex
\subsection{Supply chain attributes}
\label{sec:scattri}

Based on a topological description from literature~\cite{parhi2018topological} as well as agent attributes introduced in~\cite{bi2022model}, we describe the role of an agent in an SCN from topological and capability perspectives. 
In the following description, we will use Wheel and Rim agents from Figure~\ref{fig:egnetwork} as examples.

\subsubsection{Connectivity}

Defined as the number of in-flow and out-flow edges (i.e., transportation units) to or from agent $a_i$:
\begin{equation}
    C_i = \sum_{a_j\in V} b_{ij}+ b_{ji},
    \label{eqn:agflowdeg}
\end{equation}
where $b_{ij}=1$ if edge $(i,j)$ is associated with material or product flow from $a_i$ into $a_j$, and 0 otherwise. 
From Figure~\ref{fig:egnetwork}, the connectivity of the Wheel agent is given as $C=3+2=5$.
The \textit{Connectivity} of an agent represents the number of other agents that will be affected if this agent is disrupted.

\subsubsection{Depth}

Defined as the maximum number of edges between $a_i$ and the final layer (e.g., customer)
\begin{equation}
    D_i = \max_{a_j\in\text{Customer}}d(a_i, a_j),
    \label{eqn:agdepth}
\end{equation}
where $d(a_i, a_j)$ is the geodesic distance, defined as the minimal length of a path between two agents $a_i$ to $a_j$.
In Figure~\ref{fig:egnetwork}, assuming the store represents the customer layer, the depth of the Wheel agent is found to be 1.
The \textit{Depth} of an agent represents where the agent is located in the supply chain, thus it reflects the possible ripple effect if the agent is disrupted.

\subsubsection{Redundancy}
\begin{figure}[!t]
\centering
\includegraphics[width=1\columnwidth]{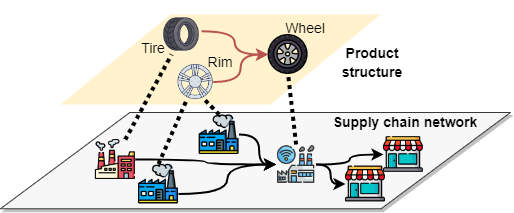}
\caption{An example of an SCN and product structure used to illustrate the proposed supply chain descriptions.}
\label{fig:egnetwork}
\end{figure}

From a capability perspective, the redundancy of $a_i$ is defined as the number of alternative agents (excluding $a_i$) that can perform the same capability as $a_i$ in the agent network:
\begin{equation}
    R_i(m) = |\{a_j|a_j\ \text{with capability}\ m, a_j\in A \setminus \{a_i\}\}|,
    \label{eqn:agcapred}
\end{equation}
where $m$ represents a specific capability of agent $a_i$. A detailed definition of $m$ is given in Section~\ref{sec:agentframwork}. 
This attribute indicates the number of backup suppliers for a given capability. In Figure~\ref{fig:egnetwork}, the network contains two agents that can produce the Rim; thus, for each Rim agent, the capability redundancy is $R_i(m)=1$.
The \textit{Redundancy} of an agent indicates whether there are backup agents to recover the product flows if this agent is disrupted.

\subsubsection{Complexity}
Defined as the sum of final product types that require products from agent $a_i$ and the material/component types necessary for $a_i$'s production:
\begin{equation}
        P_i = |\bigcup_{k\in K_i}\{k_f\ |\ k\in Z(k_f), \forall k_f\in K_f\}| + |\bigcup_{k\in K_i}Z(k)|,
    \label{eqn:agproduction}
\end{equation}
where $K_i$ represents the set of product types that $a_i$ can produce and $K_f$ represents the set of final product types in the supply chain. 
The function $Z: K\rightarrow 2^K$ maps a specific product type $k$ to the set of components and/or materials that are needed to produce $k$. 
We denote $K$ as the set of all product types in the supply chain.  
In Figure~\ref{fig:egnetwork}, the production of the wheel requires a tire and a rim, $Z(\text{Wheel})=\{\text{Tire, Rim}\}$. 
It is assumed that there is only one product type that requires the wheel. 
Thus, the production complexity of the wheel agent $P_{wheel}=1+2=3$.
The \textit{Complexity} of an agent represents the number of product types that will be affected if this agent is disrupted.

%% file: sections/3_0_approaches.tex
\section{Distributed Decision-making Using Multi-agent Framework}
\label{sec:approaches}


\input{sections/3_2_0_distributed}

%% file: sections/3_2_0_distributed.tex

\begin{figure*}[!t]
\centering
\includegraphics[width=2\columnwidth]{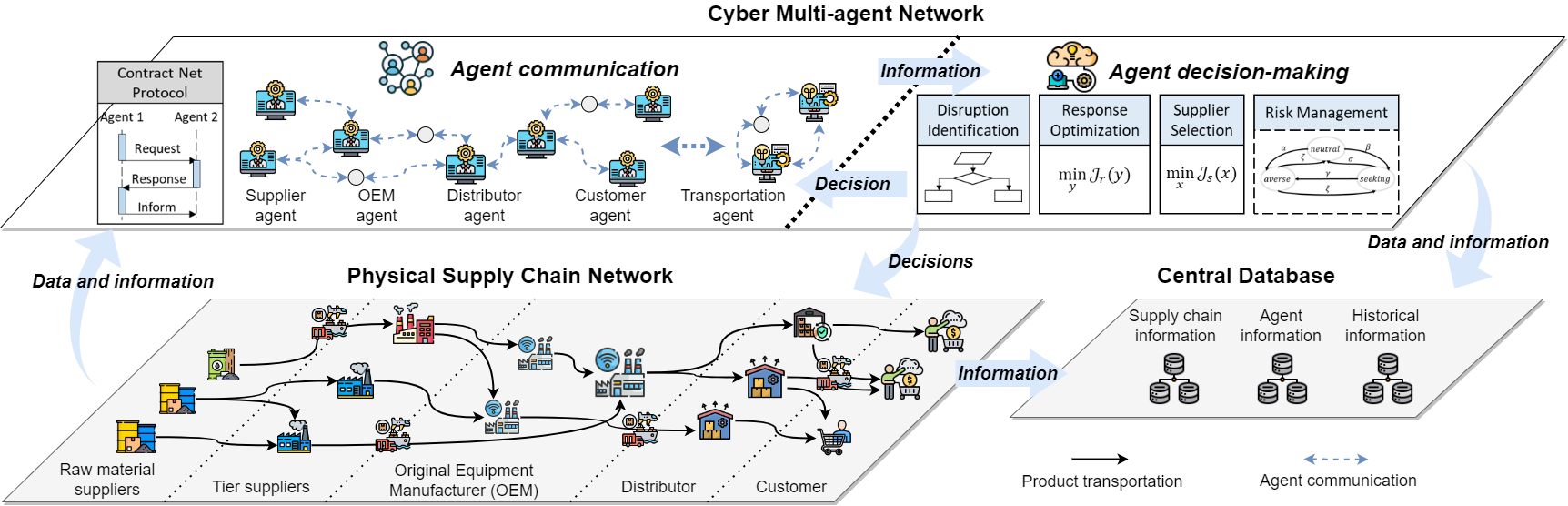}
\caption{The proposed multi-agent framework for supply chain disruption mitigation, revised based on our prior work~\cite{bi2022model}.}
\label{fig:multiagent}
\end{figure*}
As mentioned in Section~\ref{sec:intro}, centralized approaches face challenges when dealing with large and complex SCNs, as they require information from the entire network. 
In light of this, we investigate a multi-agent framework to deploy distributed decision-making for supply chain disruption mitigation using local communication.
Figure~\ref{fig:multiagent} provides a high-level overview of the proposed multi-agent supply chain framework, including a physical SCN, a cyber multi-agent network, and a central database. 
The physical SCN contains the business entities and product flows from the supply chain in the real world. 
The multi-agent cyber-network consists of an agent communication layer and an agent decision-making layer. 
Each agent is a cyber representation of its physical counterpart and will be initialized with its own version of agent architecture, as shown in Figure~\ref{fig:agent}. 
The agents obtain information from the physical supply chain and communicate with each other to share the information. 
Based on their own knowledge and shared information, agents are able to make their own decisions, such as supplier selection, and command the decided changes to the corresponding physical entity, as shown in Figure~\ref{fig:multiagent}. 
The central database stores all the information from the physical supply chain and cyber agent networks. 
In this section, we provide our design of a supply chain agent architecture and describe how agents communicate and make decisions for disruption mitigation.

\input{sections/3_2_1_agentframework}

\begin{figure}[!t]
\centering
\includegraphics[width=\columnwidth]{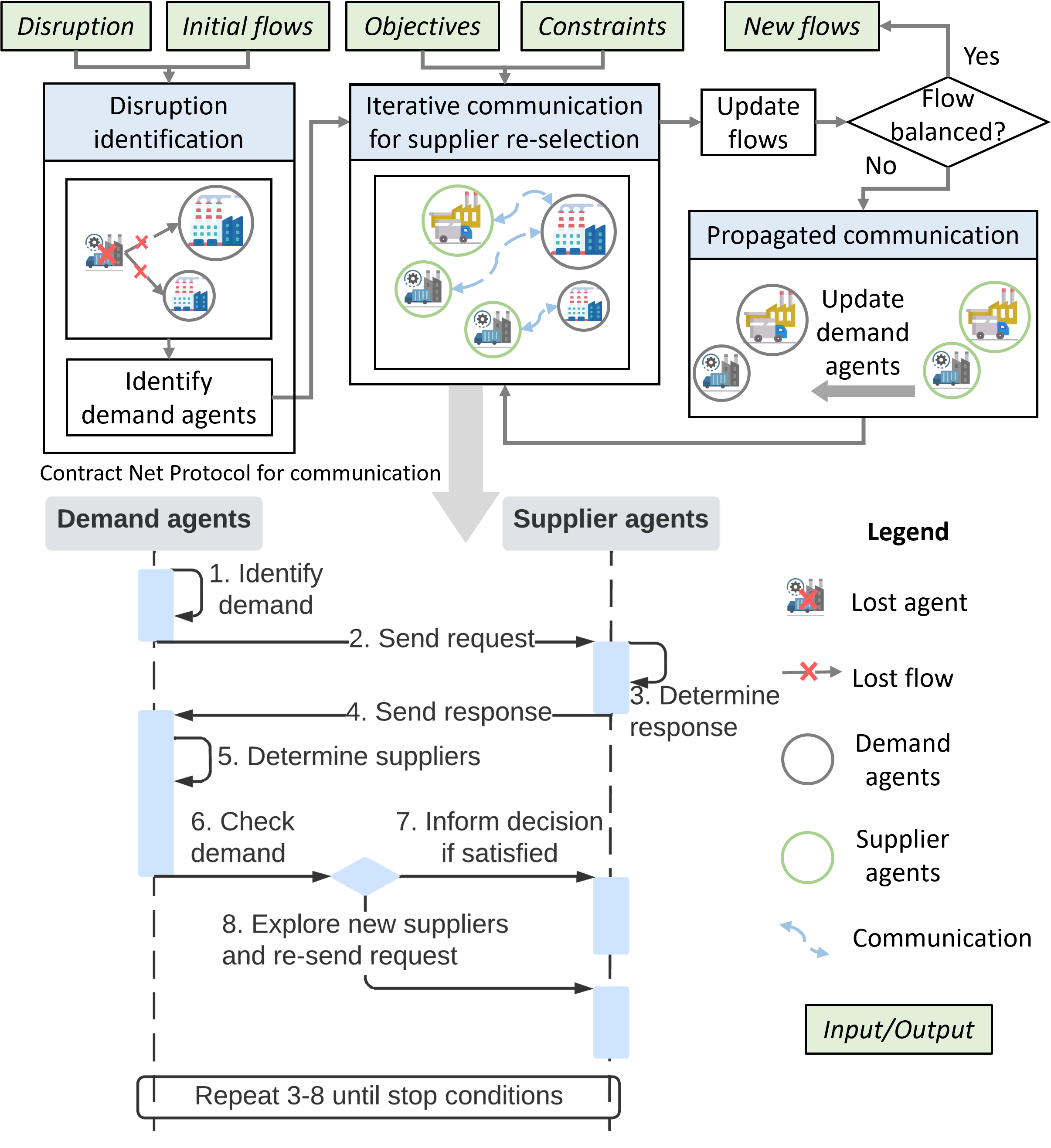}
\caption{The flow chart of the proposed agent communication and decision-making process}
\label{fig:iteration}
\end{figure}

\input{sections/3_2_2_0_communication}

Overall, the proposed model-based agent knowledge provides heuristics to guide agent communication.
In this way, agents are able to selectively communicate with other agents who possess relevant abilities to handle the disruption, thereby reducing unnecessary communication.
Moreover, the integration of agent exploration and iterative communication enables agents to thoroughly search for all potential solutions within the network.
However, it is important to note that the optimality of this approach depends on how much the users allow agents to explore and communicate.
There is a cost trade-off between exploration, and hence increased communication requirements, and performance.

%% file: sections/3_2_1_agentframework.tex
\subsection{Supply chain agent architecture}
\label{sec:agentframwork}

\begin{figure}[!t]
\centering
\includegraphics[width=1\columnwidth]{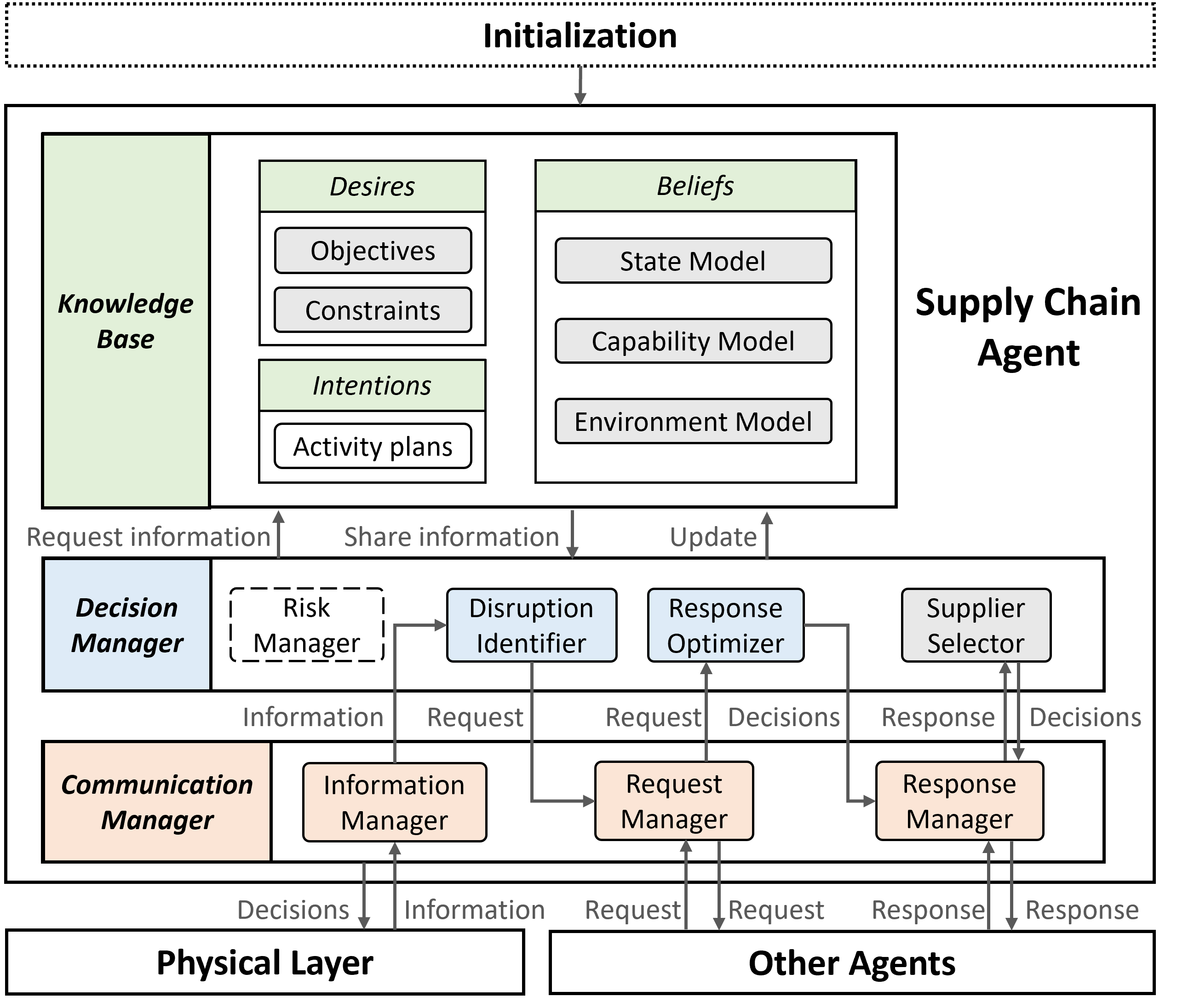}
\caption{The proposed supply chain agent architecture. The blocks in grey are examined in~\cite{bi2022model}.}
\label{fig:agent}
\end{figure}

The proposed supply chain agent architecture consists of three modules: a Knowledge Base, a Decision Manager, and a Communication Manager, following the principles of our previous architecture for manufacturing agents~\cite{kovalenko2019model, bi2023dynamic}. 
Figure~\ref{fig:agent} depicts a detailed design of the proposed agent architecture, including specific components in the modules and component-to-component information exchange. 

\subsubsection{Knowledge Base}
The belief-desire-intention (BDI) architecture has been widely used to provide a modular framework to design intelligent agents~\cite{howden2001jack}. 
Based on our prior agent models in~\cite{bi2022model}, we reformulate the agent Knowledge Base following the BDI design. 
We define the prior agent models as Beliefs and add Desires and Intentions into the Knowledge Base in this paper.

\paragraph{Beliefs} Beliefs represent what the agent knows about itself and its environment. 
Building upon our prior work~\cite{bi2022model}, the beliefs of an agent are comprised of the state, capability, and environment models. 
These models are dynamically updated (i.e., extended, shrunk, and revised) as the agent and its environment change.

\textit{State model:} The agent dynamics are described by a flow balance of varying input and output products. 
The state model describes the dynamics in terms of flow, production, and inventory based on the flow balance equation~\cite{bi2022model}:
\begin{equation}
    I_{t+1} = I_{t} + u_{t} - z_{t} + h_{t}(I_{t},u_{t}).
    \label{eq:statedynamics}
\end{equation}
Note that each variable in the state model is a vector indexed by the product types that agent $a_i$ needs or produces, where state vector $I_{t}$ represents the number of products held in the agent at time $t$; input vector $u_{t}$ represents product flows into the agent; output vector $z_{t}$ represents the product flows out of the agent; and production function $h_{t}(I_{t},u_{t})$ defines the number of used components and produced new products if the agent has production capability.

\textit{Capability model:} Describes the operational behaviors that an agent can perform in the SCN. 
The capability model consists of capability knowledge and several associated mapping functions. 
The capability knowledge is a set of capabilities $M_c=\{m_0,m_1,...\}$. 
Each capability $m_j=(o, k)$ is a tuple, where $o$ is one of the operational behaviors, including production, inventory, or transportation, and $k$ is a product type. 
Along with this high-level capability knowledge, several mapping functions are used to describe the characteristics of a capability, such as cost and capacity.

\textit{Environment model:} An agent's knowledge of other agents in the network is encapsulated in the environment model. 
Based on the agent network $G^a=(A, L)$, we can describe the local communication network for a single agent $a_i\in A$ as $G^a_i = (A_i, L_i)$, where $A_i \subseteq A$ is a subset of agents that $a_i$ can communicate with via links in $L_i$. 
These agents are grouped into several subsets based on their relationship with agent $a_i$ and stored in the environment model, denoted as $M_e=\{U_i: K\rightarrow2^{A_i}, D_i, S_i,...\}$. 
$U_i$ maps product types to a set of upstream agents from which $a_i$ can obtain the products. 
Similarly, the mapping functions, $D_i$ and $S_i$, identify the agents that are downstream agents of $a_i$ and agents that have the same capability as $a_i$, respectively. 
In this way, an agent is capable of identifying the subset of agents it needs to communicate with and exchange information in the network.

\paragraph{Desires} Desires represent the goals and requirements of an agent.
In this paper, the desires include the objective functions and the constraints for the decision-making of the agents. 
For example, an agent makes decisions to select suppliers with minimal cost (objective $\mathcal{J}$) considering the limit of the number of suppliers (constraint $\mathcal{C}$).

\paragraph{Intentions} Intentions represent the plans an agent has committed to executing. 
The intentions of an agent depend on its capabilities. 
For example, the intentions of a transportation agent include product flows it has committed to transporting, while for a supplier agent, the intentions describe the production and out-flow of products to downstream agents.

\subsubsection{Decision Manager}
The Decision Manager of an agent consists of multiple decision-making models. 
In this paper, we primarily discuss the decision-making for disruption identification, response optimization, and supplier selection.

\paragraph{Disruption Identifier} Component that identifies the consequences of a disruption, such as the lost production and flow streams.

\paragraph{Response Optimizer} Component that determines how an agent responds to other agent requests by solving an optimization model based on the response agent's objectives and constraints.

\paragraph{Supplier Selector} Component that determines how an agent selects suppliers to satisfy its demand by solving an optimization model based on the selector agent's objectives and constraints.

\subsubsection{Communication Manager}
Provides the interface for the agent to exchange information with its physical entity and other cyber agents. 
In this paper, components of the communication manager include an information manager, a request manager, and a response manager.

\paragraph{Information Manager} Collects information from and sends decisions to the agent's associated physical entity.

\paragraph{Request Manager} Sends requests from the Decision Manager to other agents and passes requests received from other agents to the Decision Manager.

\paragraph{Response Manager} Sends responses from the Decision Manager to other agents and passes responses received from other agents to the Decision Manager.

%% file: sections/3_2_2_0_communication.tex
\subsection{Agent communication and decision-making for disruption mitigation}
\label{sec:communication}

In this section, we describe the proposed agent communication strategy for disruption mitigation, including three processes: disruption identification, iterative communication for supplier re-selection, and propagated communication. 
Figure~\ref{fig:iteration} shows the overall flow of the agent decision-making and communication protocol for supplier selection. 
Algorithm~\ref{alg:overall} describes the detailed overall decision-making processes while Algorithm~\ref{alg:agcom} describes how agents iteratively communicate for supplier re-selection. 
We assume that agents can communicate regardless of the disruption and that communication links can be established or removed based on the decision-making strategy. 
Agent communication follows Contract Net Protocol (CNP)~\cite{smith1980contract}, which aligns with the Foundation for Intelligent Physical Agents (FIPA) standards~\cite{poslad2007specifying}.

\subsubsection{Disruption identification}
\label{sec:disruptionidentify}

\input{sections/3_2_2_1_identify}

\subsubsection{Iterative communication for supplier re-selection}
\label{sec:agentcommunication}

\input{sections/3_2_2_2_iterativecommunication}

\subsubsection{Propagated communication}
\label{sec:agentpropagation}

\input{sections/3_2_2_3_propagate}

%% file: sections/3_2_2_1_identify.tex
We focus on the disruption of an agent loss, which leads to losses of production and/or transportation flows. 
By periodically obtaining data and information from the physical entities, agents are able to detect disruptions that occur in their associated physical entities. 
Once a disruption occurs, the disrupted agent, denoted by $a_e$, checks its knowledge base to identify the lost flow $y_e^0$ related to it, as described in Algorithm~\ref{alg:overall}, lines 1-3. 
Based on the lost flow, $a_e$ initiates communication with downstream agents, defined as demand agents $A_{dm}$, to inform them of the disruption to their incoming production flow streams. 
In addition, the $a_e$ will inform upstream agents that provide products to the disrupted agent that the flow streams will be disrupted and may be reassigned to alternative agents. 
Since the flow streams $f_r$ are no longer balanced and meeting the requested need, the demand agents $A_{dm}$ must now find supplier agents that are capable of supplementing for the lost production of $a_e$.

\input{algorithms/algorithm1}

%% file: algorithms/algorithm1.tex
\begin{algorithm}[t]
    \caption{Distributed decision-making for disruption mitigation}
    \label{alg:overall}
    \begin{algorithmic}[1]
    \REQUIRE $a_e, f_0, \mathcal{C},\mathcal{J}$
    \ENSURE  $f_r$\\
    
    \COMMENT{\ Disruption and demand identification}
    \STATE $y_e^0\gets(y_{ejk}^0, y_{iek}^0)_{i,j\in V, k\in K}\subset f_0$\ \COMMENT{\ Identify lost flows}
    \STATE $A_{dm}\gets\{{a_j}, \forall j\in y_{ejk}^0\in y_e^0$\}\ \COMMENT{\ Identify demand agents}
    \STATE $f_r\gets f_0\ \backslash\ y_e^0$ \COMMENT{\ Get remaining flows}
    
    \COMMENT{\ Agent communication}
    \WHILE{$A_{dm}\neq \emptyset$}
    
    \item[] \COMMENT{\ Generate and update new product flows}
    
    \STATE $\Delta f_r\gets Algorithm~\ref{alg:agcom}$\\
    \STATE $f_r\gets f_r\cup\Delta f_r$

    \COMMENT{\ Check flow balance and identify new demand agents}
    \FOR{$y'_{zjk}\in\Delta f_r$}
    \IF{$a_z$ needs materials/components}
    \STATE Add $a_z$ to $A_{dm}$
    \ENDIF
    \ENDFOR
    \ENDWHILE
    \RETURN $f_r$
    \end{algorithmic}
\end{algorithm}

    
    

%% file: sections/3_2_2_2_iterativecommunication.tex
Once the disrupted agent, $a_e$, identifies the lost flow streams and informs the downstream demand agents about this disruption, the demand agents must initiate communication with alternative suppliers. 
The contract net protocol (CNP) segment of the communication process illustrated in Figure~\ref{fig:iteration} and Algorithm~\ref{alg:agcom} describes the iterative communication process that consists of four key steps in each iteration: ``Identifying needs", ``Requesting help", ``Responding to the requests", and ``Informing agents of accepted flow". 
The outcome of this process is the selection of alternative suppliers to provide the necessary product flow that has been disrupted by the loss of an agent. 
\input{algorithms/algorithm3}

\paragraph{Identify needs}
Based on the disrupted flow, a demand agent $a_j\in A_{dm}$ identifies its current need, $d_{jk} = y_{ejk}^0,\forall k\in K$. Using its knowledge base, a demand agent then retrieves the set of objectives, $\mathcal{J}_{jk}$, and constraints, $\mathcal{C}_{jk}$, (e.g. budget, delivery date) to direct the decision-making process for a new supplier.


\paragraph{Request}
Meanwhile, each demand agent $a_j$ retrieves the environment models in its knowledge base and identifies the agents where it will send a request for additional flow. 
We define $\mathcal{M}_j(k)$ as a set of agents that $a_j$ can request for product $k$. 
The demand agent $a_j$ sends out requests, denoted by $Req=(d_{jk}, \mathcal{C}_{jk})$, to all agents in $\mathcal{M}_j(k), \forall k\in K$. 
Note that all of the demand agents may send requests in parallel.

\paragraph{Response}
Request agents ($a_z\in \cup_{a_j\in A_{dm}}\mathcal{M}_j(k)$), those that receive the request from the demand agents, will check their knowledge bases and determine their ability to provide the product flows (i.e., satisfy the requests) based on their objectives $\mathcal{J}_z$ and constraints $\mathcal{C}_z$. 
They will then formulate a response that includes available product flows, $\bar{y}_z=[\bar{y}_{zjk}, a_j\in A_{dm}, k\in K]^{\mathsf{T}}$, where $\bar{y}_{zjk}$ represents the maximum units of product $k$ that $a_z$ can provide to $a_j$, and related information $\mathcal{F}_{zjk}$ (e.g., product cost). 
The request agents $a_z$ determine their responses by solving a local (individual agent) optimization model.
An example is given below:
\begin{subequations}
\label{eq:maxout}
\begin{align}
\max_{\bar{y}_z}\quad & \mathcal{J}_{z}(\bar{y}_z) = \sum_{a_j\in A_{dm}, k\in K}r_{zjk}\bar{y}_{zjk} \label{eq:maxoutobj}\\
\text{s.t.}\quad 
&\sum_{k \in K}\bar{y}_{zjk} \leq q_{zj}, \ \forall a_j\in A_{dm} \label{eqc:cap_z}\\
& \sum_{a_j\in A_{dm}}\sum_{k \in K}\bar{y}_{zjk} \leq \bar{p}_{z}, \label{eqc:cap_p}\\
& \bar{y}_{zjk}\leq d_{jk}, \forall a_j\in A_{dm}, k \in K, \label{eqc:reqconstr}
\end{align}
\end{subequations}
where~\eqref{eq:maxoutobj} maximizes agent $a_z$'s revenue, subject to the constraints of flow capacity~\eqref{eqc:cap_z} and production capacity~\eqref{eqc:cap_p} from the agent itself and the constraints of from the requested demands~\eqref{eqc:reqconstr}.
Note that agents can have their own specific objectives (e.g., inventory level and revenue) and constraints.
All requested agents send their responses, i.e., $Res=(\bar{y}_{zjk}, \mathcal{F}_{zjk})$, back to the demand agents.

\paragraph{Inform}
After receiving responses from all of the requested agents, each demand agent, $a_j$, determines the new product flow streams $y'_{j}=[y'_{zjk}, a_z\in Z_{j}(k), k\in K]^{\mathsf{T}}$ by solving another optimization.
An example is shown below:

\begin{subequations}
\small
\label{eq:distribution}
\begin{align}
\min_{y'_j}\quad & \mathcal{J}_j(y'_j, \mathcal{F}_{zjk}) = \sum_{a_z\in Z_{j}(k), k \in K} r_{zjk}y'_{zjk} + \sum_{k \in K}\rho_{jk}^d \Delta_{jk}^d \label{eq:distributionobj}\\
\text{s.t.}\quad 
&y'_{zjk} \leq \bar{y}_{zjk}, \ \forall z \in \mathcal{M}_j(k), k \in K,  \label{eq:distributioncap1}\\
& \Delta_{jk}^d \geq \sum_{a_z\in Z_{j}(k)}y'_{zjk}- d_{jk},\ \forall k \in K \label{eqc:pnlty_d}\\
&\text{other agent constraints}, \label{eq:distributioncap2}
\end{align}
\end{subequations}
where~\eqref{eq:distributionobj} minimizes the total costs and demand dissatisfaction of the demand agent. 
Constraint~\eqref{eq:distributioncap1}  denotes that the chosen new product flows cannot exceed the suppliers' responses.
Constraint~\eqref{eqc:pnlty_d} calculates the unmet demands.
Constraint~\eqref{eq:distributioncap2} presents other constraints for supplier selection, such as a limited number of suppliers for the demand agent.
Note that demand agents may have their own specific objectives and constraints.
Then the demand agents check whether their decisions can satisfy their needs with an acceptable threshold. 
If so, the demand agents inform the chosen agents to provide new product flows.


\paragraph{Iterative communication}
The above request-response-inform process takes $|A_{dm}|+|\cup_{a_j\in A_{dm}}\mathcal{M}_j(k)| + |A_{dm}|$ computations.
However, if there are agents whose needs cannot be satisfied, they will explore the environment to identify other agents that can provide the needed products (i.e., identify new suppliers within $\mathcal{M}_j(k)$). 
These agents will then repeat the request-response-inform process to determine new product flows. 
The iteration process will stop once the demand agents have identified suppliers that can meet their needs, or it has been determined that there are no suitable agents capable of providing the needs. 
In such a case, demand agents have to reduce their production due to material shortage, and inform their downstream agents that they cannot provide all of the desired products.
In this paper, this product shortage will lead to unmet demand for customers.
Note that different communication strategies could also be designed to further investigate the capability and capacity of the network to fulfill demands.
To guarantee convergence, we set an upper bound $n_e$ for the number of times that demand agents can explore.
Therefore, the complexity of Algorithm~\ref{alg:agcom} is $\mathcal{O}(n_e\times \max\{|A_{dm}|, |\cup_{a_j\in A_{dm}}\mathcal{M}_j(k)|\})$.
Note that $|A_{dm}|$ is related to the \textit{Connectivity} of the disrupted agent and $|\cup_{a_j\in A_{dm}}\mathcal{M}_j(k)|$ is related to the \textit{Connectivity} and \textit{Complexity} of the disrupted agent.
Therefore, if the disrupted agent has higher \textit{Connectivity} and \textit{Complexity}, the re-planning process generally requires more computations.

%% file: algorithms/algorithm3.tex
\begin{algorithm}[t]
    \caption{Supplier re-selection via agent communication}
    \label{alg:agcom}
    \begin{algorithmic}[1]
    \REQUIRE $A_{dm}, y_e^0, \mathcal{C},\mathcal{J}$
    \ENSURE  $\Delta f_r$\\
    \COMMENT{\ \textbf{Objective:} Find product flows to satisfy demand agents}
    \WHILE{$A_{dm}\neq \emptyset$}
    
    \item[] \COMMENT{\ Request (all the demand agents)}
    \FOR{$a_j\in A_{dm}$}
    \item[] \COMMENT{\ Identify the need of each demand agent}
    \STATE $d_{jk}\gets{y_{ejk}^0},\forall k\in K$
    \item[] \COMMENT{\ Identify agents to request}
    \STATE $a_j$ explores environments
    \STATE $\mathcal{M}_j(k)\gets$ Environment model, $\forall d_{jk}$ 
    \STATE $a_j$ requests $\mathcal{M}_j(k)$ for product need $d_{jk}$
    \ENDFOR
    
    \COMMENT{\ Response (all the agents being requested)}
    \FOR{$a_z\in \cup_{a_j\in A_{dm}}\mathcal{M}_j(k)$}
    \STATE $\Bar{y}_{zjk}, y'_{z*k} \gets \min \mathcal{J}_z$
    \STATE $a_z$ sends response $\Bar{y}_{zjk}$ to $a_j$ 
    \ENDFOR
    
    \COMMENT{\ Determine product flows and check need}
    \FOR{$a_j\in A_{dm}$}
    \STATE $y'_{zjk}\gets \min \mathcal{J}_j$
    \STATE $d_{jk} \gets d_{jk} - \sum_{a_z} y'_{zjk}$
    \IF{$d_{jk}<\epsilon$}
    \STATE $\Delta f_r$ appends $y'_{zjk}$
    \STATE $A_{dm}\gets A_{dm} \backslash\ a_j$
    \ENDIF
    
    \ENDFOR
    
    \ENDWHILE
    \RETURN $\Delta f_r$
    \end{algorithmic}
\end{algorithm}

%% file: sections/3_2_2_3_propagate.tex
The iterative communication process occurs between the demand agents and their immediate supplier agents. 
However, this process may propagate through the entire network if the process of meeting the demands of certain agents introduces new needs from the suppliers meeting those demands. 
In this manner, the suppliers become new \textit{demand} agents, resulting in a continuation of this process, as described in Algorithm~\ref{alg:overall}, lines 4-12.

\paragraph{State update}
Algorithm~\ref{alg:agcom} identifies several new product flow streams that are necessary to satisfy the needs of the initial demand agents. 
Once these flow streams are identified, the demand agents and selected supplier agents must update their states, resulting in a change to the network flow states: $f_r\gets{f_r\cup\Delta f}$. 
At this point, based on the objective of Algorithm~\ref{alg:agcom}, it is assumed that the demand agents have reached a balanced flow, while the selected supplier agents may need additional components in order to meet their new flow demands. 

\paragraph{Propagation}
Since each selected supplier agent, $a_z$, commits to providing products to meet the needs of the demand agents, this may introduce additional product/component needs from the suppliers to ensure sufficient products to meet these new commitments. 
In this case, the supplier agents no longer have balanced flow streams and must propagate demand requests in order to meet their needs to \textit{their} related supplier agents. 
The communication process of Algorithm~\ref{alg:agcom} will now be repeated; however, in this iteration, the selected supplier agents have become new demand agents. 
The propagation process stops when all of the agents have met their additional needs (e.g., the requests have been propagated through all upstream agents in the network). 
Therefore, the number of times that Algorithm~\ref{alg:agcom} is repeated depends on the \textit{Depth} of the disrupted agent ($D_d$).
The worst case is that the communication propagates to the most upstream agent, whose depth is $D_{max}=\max_{a_i\in V}\{D_i\}$.
In this case, Algorithm~\ref{alg:agcom} is repeated $\Delta D=D_{max}-D_d$ times.
Therefore, the complexity of Algorithm~\ref{alg:overall} is $\mathcal{O}(\Delta D\times \mathcal{N}_{max})$, where $\mathcal{N}_{max}$ represents the maximum number of computations in the repeated Algorithm~\ref{alg:agcom}.
Note that if the disrupted agent has higher \textit{Depth}, the re-planning process requires less computations.

%% file: sections/4_0_setup.tex
\section{Case Study set-up}
\label{sec:setup}
To conduct numerical studies to evaluate the proposed approaches, we designed an example based on a supply chain for vehicle cockpits. 
In this section, we describe the supply chain instance, introduce several disruption scenarios, and derive metrics for performance evaluation.

\input{sections/4_1_network}

\input{sections/4_2_scenario}

\input{sections/4_3_metrics}

%% file: sections/4_1_network.tex
\subsection{Supply chain instance}
\label{sec:network}
\subsubsection{Product structure}
\begin{figure}[!t]
\centering
\includegraphics[width=\columnwidth]{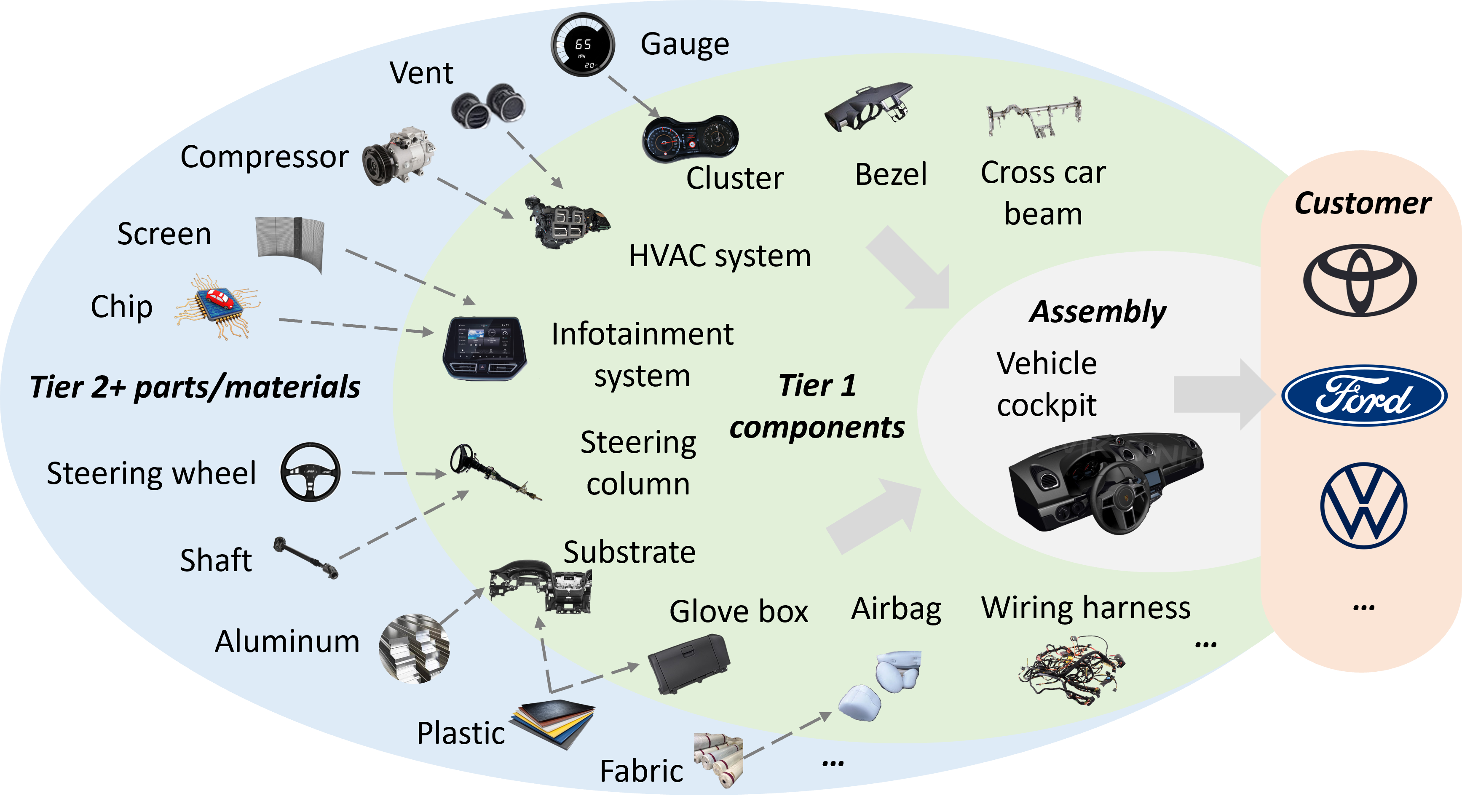}
\caption{The simplified product structure for automotive cockpit}
\label{fig:cockpit}
\end{figure}
We consider a vehicle cockpit supply chain, consisting of cockpit assembly plants, their customers (i.e., vehicle assembly plants), and their suppliers for components and materials. 
Here we summarize the supply chain product structure and network. 
Figure~\ref{fig:instance} shows that in our example vehicle cockpits represent the final product and are assembled using several manufactured components, comprised of different parts and/or materials. 

In this instance, there are 3 different models of vehicles, and each model has demands for 1, 2, or 3 styles of cockpits. 
Between auto and cockpit assembly plants, we have the following assumptions:
\begin{itemize}
    \item Each auto assembly plant only makes 1 type of vehicle model;
    \item Each cockpit assembly plant can produce multiple styles of cockpits.
\end{itemize}

Each cockpit requires 10 components to be assembled, as shown in the green ellipse in Figure~\ref{fig:cockpit}, but different styles may need different component types. 
Between cockpit assembly and component suppliers, we have the following assumptions:
\begin{itemize}
    \item The cockpits for the same auto model require the same type of cluster, substrate, glove box, HVAC system, cross-car beam, and steering column;
    \item Each style of cockpit requires a unique type of infotainment system, wiring harness, and a combination of different bezel types;
    \item All the cockpits use the same type of airbag.
\end{itemize}

Each component needs certain types and amounts of parts and/or materials in order to be produced. 
The blue ellipse in Figure~\ref{fig:cockpit} provides examples of some of the parts/materials needed for these components. 
Between component suppliers and part/material suppliers, we have the following assumptions:
\begin{itemize}
    \item Different types of components may require different part/material types, yet they may share the same part/material suppliers;
    \item The part/material suppliers represent the most upstream suppliers in this instance.
\end{itemize}

\subsubsection{Supply chain network}

\begin{figure}[!t]
\centering
\includegraphics[width=\columnwidth]{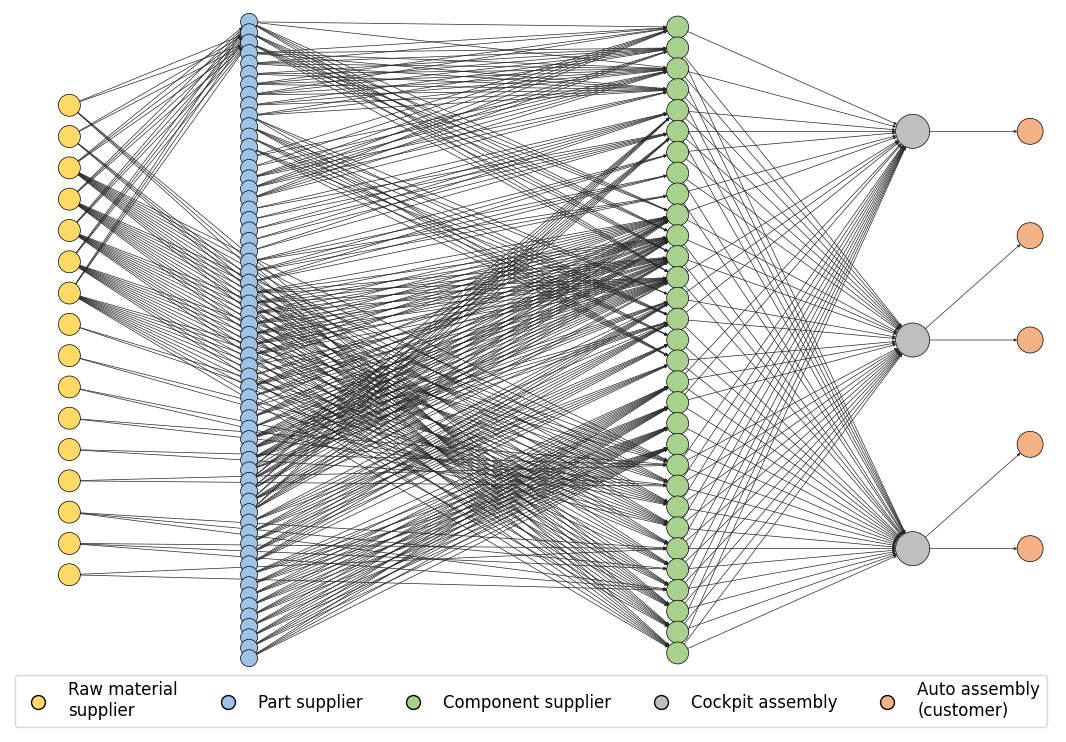}
\caption{The SCN instance used for case studies.}
\label{fig:instance}
\end{figure}

Based on the product structure above, we designed an SCN with 117 supplier/customer agents and 413 transportation agents, as shown in Figure~\ref{fig:instance}. 
These agents correspond to \textit{physical} elements within the supply chain and are equipped with the proposed agent architecture to conduct communication and decision-making. 
The network instance contains 5 customers, 3 cockpit assembly plants, 31 component suppliers, 62 part suppliers, and 16 raw material suppliers that are connected via distinct transportation units.
The 5 auto assembly plants represent customers that have placed demands for different style cockpits. 
Each cockpit assembly plant produces cockpits for a specific auto model type. 
For each type of component, part, and material, we have multiple suppliers that have production capabilities. 
Each supplier or transportation agent has its own production or transportation cost and capacity.

%% file: sections/4_2_scenario.tex
\subsection{Benchmark and disruption scenarios}
\label{sec:scenario}
As discussed in Section~\ref{sec:intro}, existing distributed approaches in the field of supply chain management typically rely either on pre-defined disruption scenarios or rule-based decision-making.
It is challenging to evaluate and compare with these approaches effectively without access to their underlying agent design, database, and implementation details.
Additionally, these methods do not focus on the problem of supply chain disruption response, thus comparing and assessing their performance accurately for this problem may be hindered, potentially leading to incomplete or misleading conclusions.
Furthermore, though there are other distributed approaches utilized in other fields, such as manufacturing rescheduling, multi-robot control, these approaches cannot be directly applied to the specific problems posed by supply chain disruption mitigation.

Therefore, to benchmark our distributed approach against a more common decision-making strategy, we evaluate the performance of both a centralized and distributed decision-making approach during various disruption scenarios.  
With visibility of all entities and their status in the SCN, the centralized model provides a highly communicative yet globally optimal reconfiguration plan. 
We use the centralized model that we developed in~\cite{bi2022model} to generate optimal initial product flows as a steady state before a disruption occurs. 
Once a disruption is identified, the model is updated by modifying the network structures, parameters, and constraints. We then run the updated model to determine newly optimized decisions about the product flow and production schedule as a response to the disruption.
Though it is not straightforward to analyze how the solver computes the solution for the centralized model, the input size of the centralized model is much larger than the proposed distributed approach.
The centralized model re-optimizes all the agents considering associate constraints while the proposed distributed approach only consider a subset of agents within the network.

The supply chain is initiated with a product flow plan derived  by solving the centralized model optimization described above. 
The supply chain descriptions defined in Section~\ref{sec:scattri} can be used to describe the role of each agent in this supply chain instance. 
Note that in the initial product flow plan, not all agents will have active production or transportation roles.

To evaluate a disruption from the loss of a single agent, we design the case study with the following rules:
\begin{itemize}
    \item The lost agent should have production tasks in the initial plan;
    \item In each scenario, only one supplier agent becomes disrupted;
    \item Agents can exhibit production and transportation capabilities of 30\% over their initially defined amount for an additional 50\% unit cost.
\end{itemize}
In the pre-determined initial plan, there are 84 agents that exhibit production tasks, thus we run 84 scenarios, starting from upstream agents to downstream agents. 
For each scenario, we implement the centralized and distributed decision-making approaches from Section~\ref{sec:approaches} to generate a new flow plan without the use of the lost agent.

The decision-making is focused on optimizing the system performance by minimizing a cost function, $\mathcal{J}$, at the network level (centralized) or local agent level (distributed):
\begin{equation} \label{eqn:caseobj}
\begin{split}
\mathcal{J}
=&\sum_{(i,j) \in E, k \in K} c_{ijk}y_{ijk} +
\sum_{i \in V, k \in K} e_{ik}p_{ik}\\
&+\sum_{i \in V, k \in K}\rho_{ik}^d \Delta_{ik}^d + \sum_{(i,j) \in E}\rho_{ij}^{E}\Delta_{ij}^E + \sum_{i \in V}\rho_i^{V}\Delta_i^V, 
\end{split}
\end{equation} 
where the first two elements represent transportation and production costs when overcapacity must be applied, and the last three represent the penalty costs for unmet demand and the addition of new agents and edges. 
Note that at the local agent level, the cost function is applied across several local agents rather than the entire network as with the centralized approach. 
As mentioned in Section~\ref{sec:approaches}, an agent explores the environment to identify agents that can meet production needs in the case of a disruption. It is assumed that agents only interact with other agents that are within their network and therefore show up in their knowledge base. 
If the existing network cannot satisfy the required demands, an agent will seek to build connections with new agents. This exploration process will trigger a penalty if new agents are added to the existing network. The proposed cost function is used to provide an example. Additional objectives can also be investigated within this framework through the selection of different elements within the cost function.

%% file: sections/4_3_metrics.tex
\subsection{Metrics for performance evaluation}
\label{sec:metrics}

In order to evaluate the impact of an agent's attributes within the network on the outcomes of different decision-making strategies, we define several key performance metrics. 




\subsubsection{Overage cost}
We define overage cost $O$ as the total cost for any transportation or production flow that exceeds the original agent capacity. 
This metric represents additional efforts by the agents to address the disruption.
\begin{equation}
    O=\sum_{(i,j) \in E, k \in K} \alpha_{ij}c_{ijk}y_{ijk}^o +
    \sum_{i \in V, k \in K} \beta_i e_{ik}p_{ik}^o,
    \label{eqn:overcapacity}
\end{equation}
where $\alpha_{ij}$ and $\beta_i$ are the multipliers for the increased cost of over-capacity flow and production; $y_{ijk}^o$ and $p_{ik}^o$ are the amount of over-capacity flow and production that are determined by the optimization.

\subsubsection{Network changes}
We define network changes $N_c$ as the sum of the number of agents that changed their existing production amount plus the number of flow channels that are changed in terms of the type and/or amount of products.
\begin{equation}
    N_c=|\{a_i|p_i\neq p_i',\forall i\in V\}| + |\{(i,j)|y_{ij}\neq y_{ij}',\forall (i,j)\in E\}|,
    \label{eqn:changed}
\end{equation}
In~\eqref{eqn:changed}, $p_i$ is the initial production of $a_i$ and $y_{ij}$ is the initial flow of edge $(i,j)$; $p_i'$ and $y_{ij}'$ are the new production and flow, respectively.

\subsubsection{Network additions} 
We define Network additions $N_a$ as the sum of the number of new agents and edges that are added to the network to address the disruption.
\begin{equation}
    N_a=\sum_{i \in V} \max{\{0, \xi'-\xi\}} + \sum_{(i,j) \in E} \max{\{0, \zeta'-\zeta\}}
    \label{eqn:added}
\end{equation}
Here $\xi$ and $\zeta$ indicate the usage of agents and edges in the initial plan (1 if used, 0 otherwise), respectively, and $\xi'$ and $\zeta'$ represent the new plan. 
Both $N_c$ and $N_a$ represent how the network changes to respond to the disruption in terms of the overall production and flow in the network. 
In practice, changing existing flow streams and production types or introducing new suppliers and transportation units may require a significant amount of business effort, and may not be practical in many instances.


\subsubsection{Agent communication}
We define agent communication $M$ as the number of communication exchanges used to determine a response to the disruption. 
The communication effort $M$ in the centralized method includes the request for re-running the model, the requests to and responses from all the agents in the supply chain to collect information, and the notifications to the agents whose flow and/or production plan need to change:
\begin{equation}
    M=1+2|V|+N_c+N_a
    \label{eqn:centralcommunication}
\end{equation}
In the distributed method, $M$ includes all the agent requests, responses, and inform messages, as defined in Section~\ref{sec:agentcommunication}.


%% file: sections/5_0_casestudy.tex
\section{Case studies results and discussion}
\label{sec:casestudy}

In this section, we present a summary of the case study results for the various scenarios and investigate how agent attributes, within the context of a specific decision-making approach, impact the different performance metrics. 
Based on these results, we provide insight for users to consider when determining an approach to use for disruption response.

\input{sections/5_1_overview}

\input{sections/5_2_performance}

\input{sections/5_3_insights}

%% file: sections/5_1_overview.tex
\subsection{Overview of the results}
\label{sec:resultoverview}


In this case study we evaluated 84 disruption scenarios. 
Within these 84 cases, there were 11 scenarios in which at least one of the approaches was unable to find a solution that could satisfy all of the customer demands. 
In these scenarios, the network exhibited a redundancy of zero, $R_i(m)=0$, or insufficient remaining capacity to recover all the production losses of the disrupted agent. 
For these cases, the centralized approach was used to meet the demand by re-optimizing the entire SCN to redistribute the capacity and production capabilities. 
For the remaining 73 scenarios, the distributed and centralized approaches were able to find new plans that satisfied all the demands. 
Importantly, the computation time for the distributed decision-making approach was 99\% faster than the centralized approach. 
The obtained results in our study provide validation for our assumption that the redundancy of the disrupted agent is indeed a necessary condition for recovering performance in SCNs. The presence of redundant agents plays a crucial role in maintaining and restoring the overall system performance in the face of disruptions.
Moreover, the proposed distributed approach demonstrates its capability to find a solution for the loss of any arbitrary agent, as long as a viable solution exists within the network.

\input{tables/metrics.tex}

A comparison of the performance of the supply chain reconfiguration as a function of the decision strategy and performance metrics is shown in Table~\ref{tab:metrics}. 
Network changes and communication use are minimized by the distributed approach, while overage costs and network additions generally result in similar performances for either the centralized or distributed approach. In the following analysis, we focus on the 73 scenarios where the production demands are satisfied.

%% file: tables/metrics.tex
\begin{table}[t]
\caption{Metric evaluation across individual scenarios}
\label{tab:metrics}
    \begin{tabularx}{\columnwidth}{lcc}
    \hline
    \multirow{3}{*}{\textbf{Metrics}} & \multicolumn{2}{c}{\textbf{Number of scenarios (73 in total) where}}\\
     & distributed approach & two approaches\\
     & performs better & are similar \\
    \hline
    
    Network changes $N_c$ & 73 & 0 \\
    Communication $M$ & 73 & 0\\
    Overage cost $O$ & 18 & 42 \\
    Network additions $N_a$ & 0 & 42  \\
    
    \hline
    \end{tabularx}
\end{table}

%% file: sections/5_2_performance.tex
\subsection{Performance}
\label{sec:performance}

In this section, we investigate how agent attributes impact the performance metrics for both centralized and distributed approaches. Note that since redundancy mainly affects demand satisfaction and depth only affects partial metrics, we focus on agent connectivity and complexity.
To illustrate how the attributes affect performance, we categorize the 73 agents into four categories, which are combinations of low and high connectivity and complexity.
The connectivity of the 73 agents goes from 1 to 11 and we choose 5 as the cutoff between low and high.
The complexity of the 73 agents goes from 1 to 15 and we choose 7 as the cutoff between low and high.
Based on these criteria, there are 56 agents with low connectivity and low complexity, 5 agents with low connectivity and high complexity, 2 agents with high connectivity and low complexity, and 10 agents with high connectivity and high complexity.

\subsubsection{Network changes}

\begin{figure}[!t]
\centering
\includegraphics[width=\columnwidth]{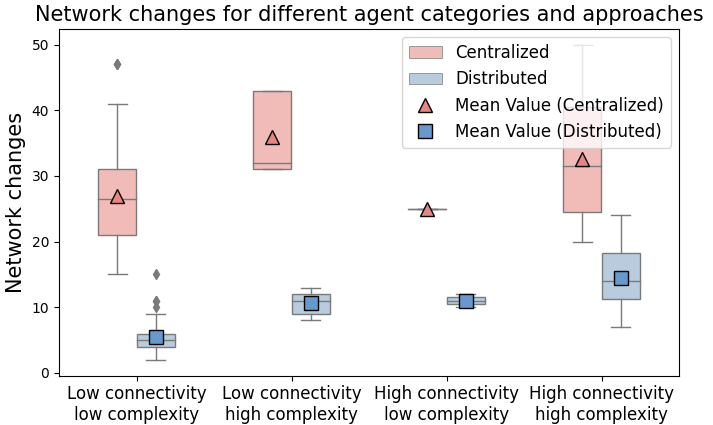}
\caption{Number of network changes for different categorized agents based on attributes using centralized and distributed approaches.}
\label{fig:metricN}
\end{figure}

As shown in Table~\ref{tab:metrics}, the distributed approach results in fewer network changes than the centralized approach in all 73 scenarios, while the exact number of network changes is related to the attributes of the lost agent. 
We categorize agents based on their connectivity and complexity, and present the outcome of these categorized scenarios, as shown in Figure~\ref{fig:metricN}. 
For the centralized approach, agent connectivity has minimal impact on network changes; however, if the lost agent has high production complexity, the centralized approach causes more network changes than the scenarios where the lost agent has low complexity. 
Since the centralized approach minimizes the total objective without considering how it will change the production and transportation of individual agents, the results show that the capability attribute (i.e., complexity) has more effect on network changes in the centralized approach than topological attributes (i.e., connectivity).

For the distributed approach, we observe that agents with high connectivity in the network (topology perspective) or high product complexity (capability perspective) result in more changes to the network. 
Specifically, Figure~\ref{fig:metricN} shows that the number of network changes increases as either connectivity or complexity become higher. 
These two attributes have a similar effect on the network changes for the distributed approach since they both determine whether the agent needs to propagate its local negotiation to other agents, thus leading to additional network changes.

In addition, the difference in network changes between the centralized and distributed approaches becomes smaller when the disrupted agent has high connectivity and complexity. 
This is because these agents may require communication across a large portion of the network, leading to additional network changes that mirror the quantification of changes from the centralized approach.

\textbf{Summary:} High complexity leads to more network changes for both the centralized and distributed approaches, while high connectivity only impacts the distributed approach.

\subsubsection{Communication}

\begin{figure}[!t]
\centering
\includegraphics[width=\columnwidth]{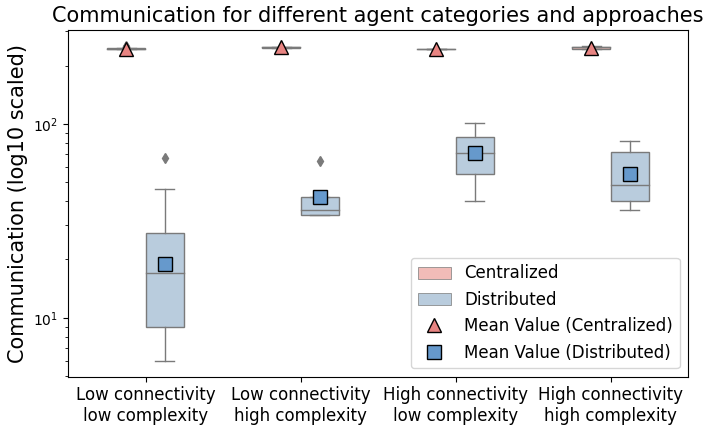}
\caption{Number of agent communication exchanges for different categorized agents based on attributes using centralized and distributed approaches.}
\label{fig:metricM}
\end{figure}
As shown in Table~\ref{tab:metrics}, the distributed approach requires less communication than the centralized approach in all 73 scenarios. 
Figure~\ref{fig:metricN} shows how agent connectivity and complexity impact communication. 
For the centralized approach, communication is not influenced by the level of connectivity and complexity of the disrupted agent. 
As defined by~\eqref{eqn:centralcommunication}, the communication for the centralized approach is dependent on the network size $|V|$.

For the distributed approach, higher connectivity or complexity leads to more agent communication. 
Similar to the performance in network changes, communication showcases the distributed approach to computing a new plan using local negotiation and only propagates communication as needed. 
However, for the agents with high connectivity, the level of complexity does not impact the communication needs for the distributed approach. This indicates that the topological attribute connectivity has more impact on communication than production complexity since it reflects the ripple effect that may go through the supply chain and lead to more communication.

The difference in communication between the centralized and distributed approaches also decreases for agents with high connectivity or complexity, since these agents may require communication through the entire network.

\textbf{Summary:} Connectivity and complexity do not impact communication requirements for the centralized approach.
High connectivity and complexity lead to more communication for the distributed approach.

\subsubsection{Overage cost}

\begin{figure}[!t]
\centering
\includegraphics[width=\columnwidth]{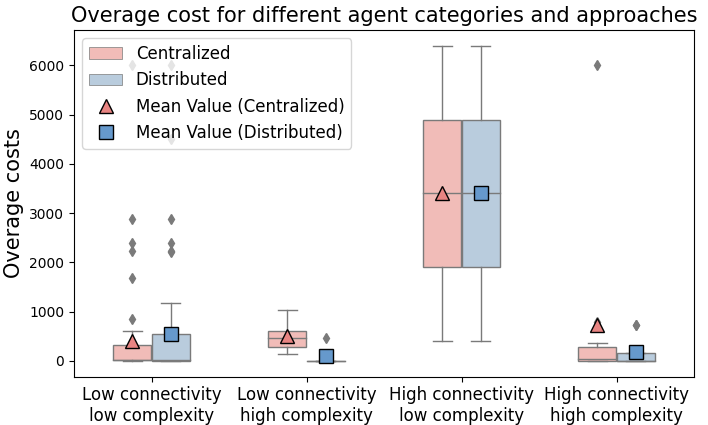}
\caption{Overage cost for different categorized agents based on attributes using centralized and distributed approaches.}
\label{fig:metricO}
\end{figure}
One might expect that a centralized approach, which optimizes flow across the entire SCN, would result in lower production costs, especially when overage costs are taken into consideration. However, Table~\ref{tab:metrics} shows that out of the total 73 scenarios, there were 18 scenarios where the distributed approach computed lower-cost solutions, and 42 scenarios where the distributed approach provided plans with similar costs to the centralized approach. 

To investigate these results, we present the overage costs based on different agent attributes, as shown in Figure~\ref{fig:metricO}. 
For both the centralized and distributed approaches, high overage costs come from low-complexity agents. 
Low complexity represents agents that require a small number of components for production or that produce variants that have limited use in the final products. Such an agent generally has a limited number of redundant suppliers with limited excess capacity. This leads to high overage costs in order to meet production needs. Overage is related to capability rather than topological attributes. 

The influence of the overage cost differs depending on the decision-making strategy. 
The distributed approach does not provide a global view, thus this approach selects several backup suppliers so that the over-capacity is low, which causes more costs through added edges. 
The centralized approach has a full network-level view, and it chooses one backup supplier with over-capacity to avoid adding additional flow channels since the centralized objective contains a large penalty for adding new agents and edges.

\textbf{Summary:} Low complexity leads to higher overage costs for both centralized and distributed approaches while connectivity has no effect.

\subsubsection{Network additions}

\begin{figure}[!t]
\centering
\includegraphics[width=\columnwidth]{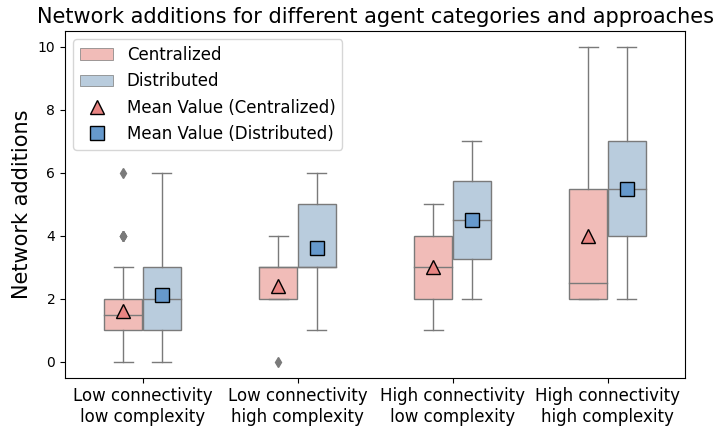}
\caption{The numbers of network additions for different categorized agents based on attributes using centralized and distributed approaches.}
\label{fig:metricH}
\end{figure}

As shown in Table~\ref{tab:metrics}, the distributed approach results in similar network additions in 42 of 73 scenarios, while the centralized approach provides fewer network additions in the other scenarios. 
Figure~\ref{fig:metricN} shows that for both the centralized and distributed approaches, the network additions increase as connectivity and complexity increase. 
Though the connectivity and complexity seem not to impact network additions significantly, it could be limited by the instance since there are not lots of unused agents to be added as a disruption response.

In addition, the distributed approach results in more network additions than the centralized approach when the lost agents have high complexity. 
As discussed above for the overage cost, the distributed approach selects several backup suppliers to lower over-capacity productions and flows but causes more added edges due to its local view of the network. 
The centralized approach utilizes its network-level view to minimize the overall additional agents and edges.

\textbf{Summary:} High connectivity or high complexity can result in more network additions for both centralized and distributed approaches.

%% file: sections/5_3_insights.tex
\subsection{Managerial insights}
\label{sec:insights}

From the results above, we can derive some insight about how agent attributes impact performance of disruption response for centralized and distributed approaches.
\subsubsection{Agent attributes}
The results show that the centralized approach's performance is more affected by agent complexity (capability attribute) than agent connectivity (topological attribute). 
This conclusion validates the study in~\cite{kim2015supply}, which states that high-connectivity agents are not necessarily critical to disruptions.
For the distributed approach, both complexity and connectivity impact the performance metrics mentioned above. 
It can be concluded that the performance of the distributed approach appears to be more sensitive to agent attributes than the centralized approach. 
Therefore, additional agent attributes should be investigated to further analyze the performance of distributed approach.
The conclusions from this work indicate, to some degree, that capability attributes are important to be considered in supply chain models, no matter what decision-making approaches to be applied.
\subsubsection{Performance evaluations}
From the performance perspective, the results above provide information for supply chain managers about how agent connectivity and complexity impact a specific set of performance metrics for both centralized and distributed approaches. 
However, in practice, enterprises and practitioners usually aim for multiple objectives. 
Though users could define any objectives to be optimized, numerical issues may occur if there are multiple objectives, e.g., hard to determine different weights. 
For this case study, in the vast majority of scenarios, the distributed approach provides solutions that have similar objective value to the centralized approach while also requiring fewer network changes and communication. 
This indicates the distributed approach may provide faster solutions that do not rely on information from the entire network at the cost of overage expenses and local optimality. 
The theoretical analysis presented in Section~\ref{sec:communication} and description of the centralized model in Section~\ref{sec:scenario} demonstrate the potential reduction in both communication and computational efforts achieved by our distributed approach. 
By leveraging local communication and the model-based agent knowledge, the proposed framework reduces extensive information exchange across the entire SCN, leading to more efficient decision-making.
However, the optimality of the distributed approach is contingent upon user-defined objectives and allowed agent exploration and iterative communications.
\subsubsection{Decision-making approaches}
The proposed distributed approach can serve as an alternative strategy in situations where centralized approaches face challenges, such as agile response requirements and high heterogeneity within the SCN.
The individual design of agents provides flexibility to manage the supply chain heterogeneously and agents' local communication enables quick responses.
However, unlike centralized approaches, the local view of agents using distributed approach may result in the potential of losing optimality.
Therefore, it is important to understand how different agent attributes impact the effectiveness and performance of both centralized and distributed approaches.
One example in the results above is that when the disrupted agent has high connectivity and complexity, the distributed approach tends towards that of the centralized approach.
In this scenario, a large amount of communication has to be used to determine a response to the disrupted agent. 
In practice, the choice of the decision-making approach largely depends on the time scale of the disruption. 
If an agent is expected to be offline for a short time, the distributed approach can give a good solution quickly, with minimal changes to the rest of the supply chain. 
For a long-term disruption, it may be worthwhile to re-run the centralized model to provide a new globally optimal plan, although a short-term modification based on the distributed approach may provide a good temporary solution.
To conclude, this work can be used to provide valuable insights for decision-makers to choose strategies depending on disruptions to enhance supply chain resilience and achieve better overall performance.
\subsubsection{Generality}
Based on the complexity analysis and the tested 84 agents in the case study, these derived insights can be generalized to some extent to other supply chains. 
However, the insights should not be interpreted as definitive inference statements.
For example, one cannot conclude that agents with higher connectivity will always result in more network changes compared to agents with lower connectivity.
More importantly, this work presents a generalized multi-agent framework that allows for investigating the correlations between agent attributes and performance. Users have the flexibility to model their own supply chains, customize metrics, and test different disruption scenarios. It is important to highlight that although our focus is on the disruption caused by the loss of an agent, this proposed approach is also applicable to disruptions related to new customer demand. In such cases, the customer itself becomes the disrupted agent as well as the demand agent, triggering the proposed agent communication strategy.
However, for other types of disruptions, such as lead time disruptions~\cite{estrada2023multi} or the introduction of new agents into the network, the current framework would need to be extended to incorporate these features. 
Further research and development would be necessary to expand the capabilities of the framework to handle such disruptions effectively.

%% file: sections/6_0_conclusion.tex
\section{Conclusion}
\label{sec:conclusion}

In this paper, we investigate the impact of network attributes on the decision-making strategy used to address supply chain disruptions. 
We first provide supply chain descriptions from a capability and topology perspective, and describe individual enterprise agents as a function of these network attributes. 
Then we reformulate our multi-agent architecture to allow agent exploration and iterative communication behaviors. 
To conduct a performance comparison, we apply a standard centralized modeling approach and our proposed distributed agent-based approach to a disrupted complex SCN. 
Through the case study, we analyze the performance of the decision-making strategies with several metrics based on the attributes of the disrupted supply chain entities in a complex SCN. 
The proposed work can be used to provide information as a decision support system to determine a decision-making strategy that optimizes user-defined performance metrics in response to supply chain disruptions. 
Future work will include developing uncertainty and risk models to incorporate a heterogeneous risk management mechanism into the agent decision-making process.

\section*{Acknowledgments}
The authors would like to thank Thomas Peretti (retired automotive professional), and John Faris (retired from Ford Motor Company) for their input in the discussions about the design of the supply chain case study.


%% file: main.bbl

%% file: bio/biotext.tex
\begin{IEEEbiography}[{\includegraphics[width=1in,height=1.25in,clip,keepaspectratio]{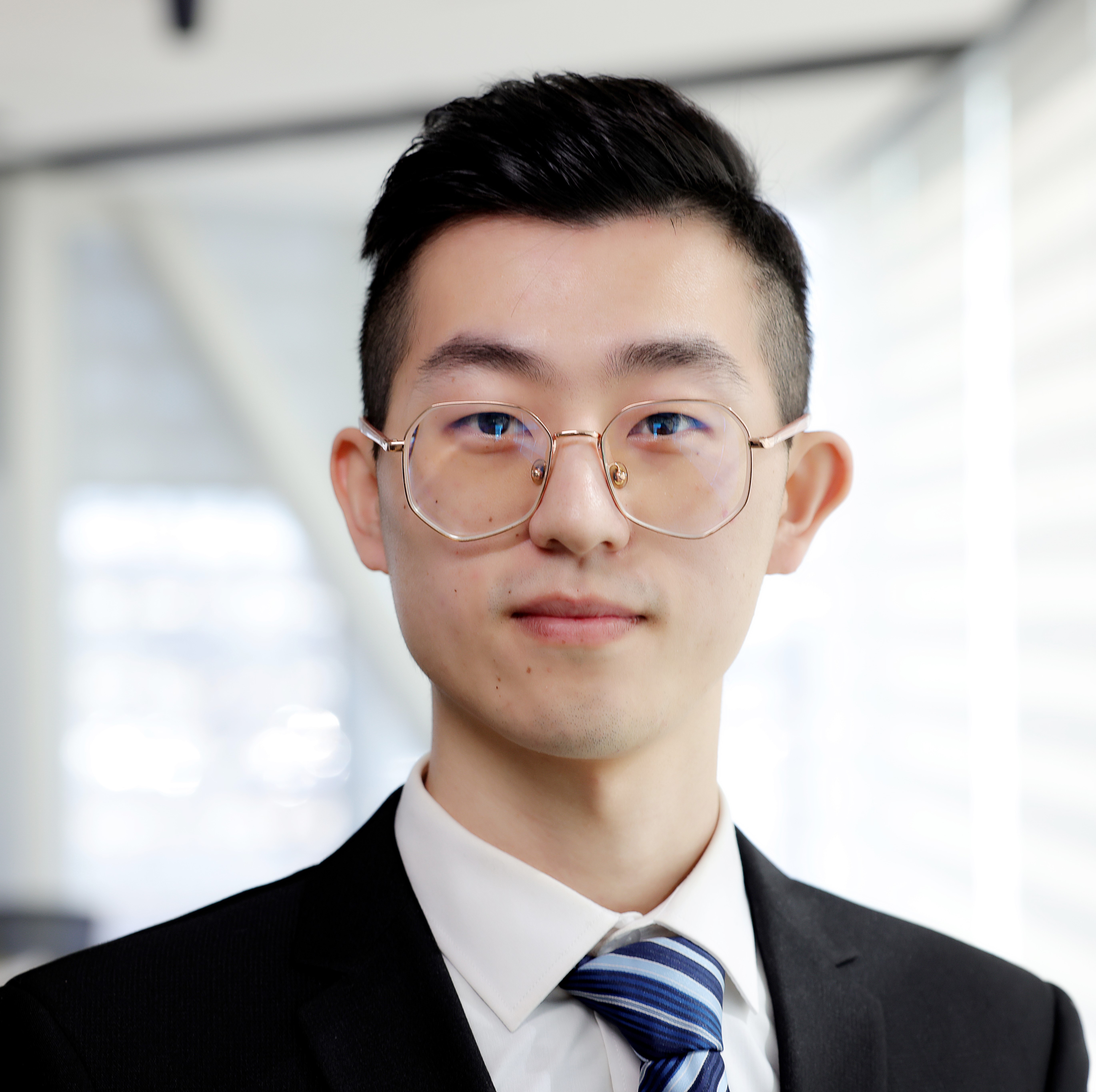}}]{Mingjie Bi}
(Student Member, IEEE) received the B.S. degree in Marine Engineering from Huazhong University of Science and Technology, China, in 2018, and the M.S. degree in Mechanical Engineering and Ph.D. degree in the Robotics from the University of Michigan, Ann Arbor, in 2020 and 2023. 
He currently works at Hitachi America Ltd. as a smart manufacturing researcher.
His research interests include system-level control, multi-agent systems, smart manufacturing, agile supply chain, and robotics.
\end{IEEEbiography}

\begin{IEEEbiography}[{\includegraphics[width=1in,height=1.25in,clip,keepaspectratio]{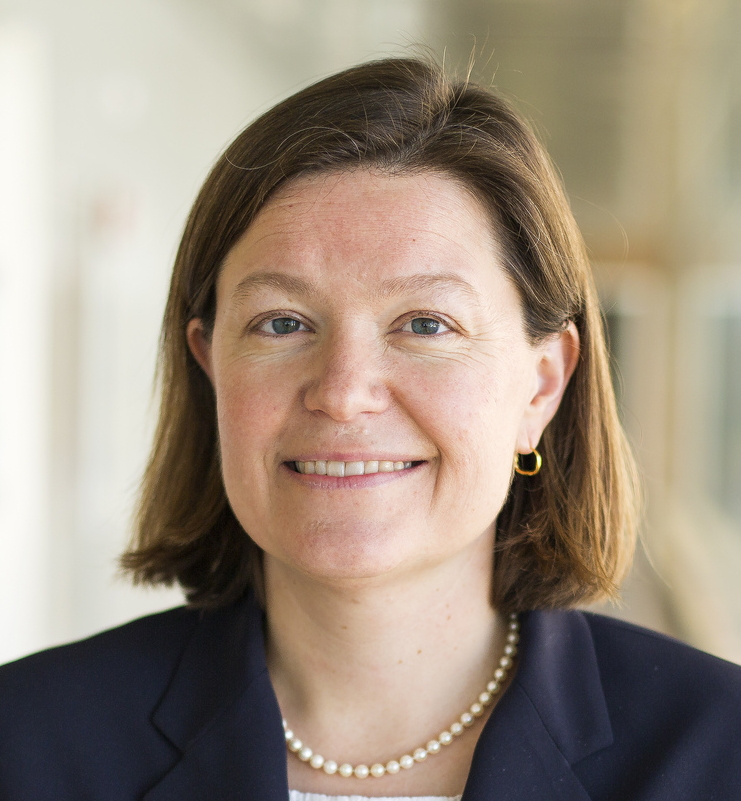}}]{Dawn M. Tilbury}
(Fellow, IEEE) is the inaugural Ronald D. and Regina C. McNeil Department Chair of Robotics at the University of Michigan, and the Herrick Professor of Engineering. She received the B.S. degree in Electrical Engineering from the University of Minnesota, and the M.S. and Ph.D. degrees in Electrical Engineering and Computer Sciences from the University of California, Berkeley.  Her research interests lie broadly in the area of control systems, including applications to robotics and manufacturing systems.  From 2017 to 2021, she was the Assistant Director for Engineering at the National Science Foundation, where she oversaw a federal budget of nearly \$1 billion annually, while maintaining her position at the University of Michigan. She has published more than 200 articles in refereed journals and conference proceedings.  She is a Fellow of IEEE, a Fellow of ASME, and a Life Member of SWE.
\end{IEEEbiography}

\begin{IEEEbiography}[{\includegraphics[width=1in,height=1.25in,clip,keepaspectratio]{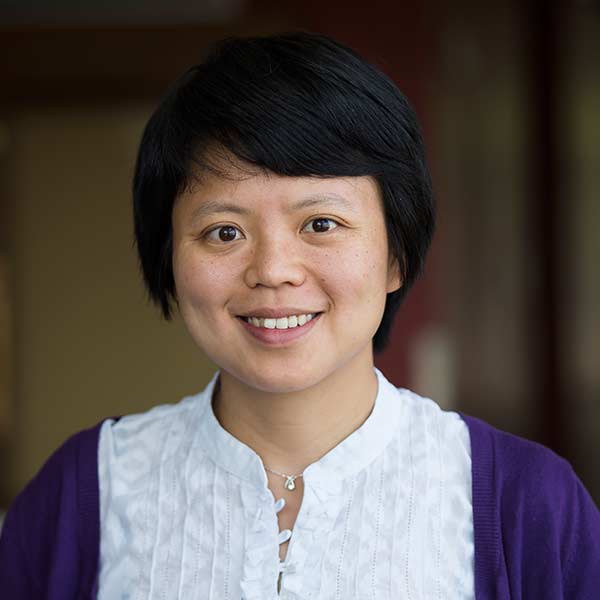}}]{Siqian Shen}
(Member, IEEE) received the B.S. degree in Industrial Engineering from Tsinghua University,  China, in 2007, and the M.S. and Ph.D. degrees in Industrial and Systems Engineering from the University of Florida, USA, in 2009 and 2011, respectively. She is a Professor in the Department of Industrial and Operations Engineering at the University of Michigan, with joint appointment in the Department of Civil and Environmental Engineering. She served as an Associate Director for the Michigan Institute for Computational Discovery \& Engineering (MICDE) from 2016-2023. 
Her research interests include stochastic programming, network optimization, and integer programming, with applications in transportation, logistics, and energy systems. She was a recipient of the Department of Energy (DoE) Early Career Award, IBM Smarter Planet Innovation Faculty Award, and the 1st place of the IIE Pritsker Doctoral Dissertation Award. 
\end{IEEEbiography}

\begin{IEEEbiography}[{\includegraphics[width=1in,height=1.25in,clip,keepaspectratio]{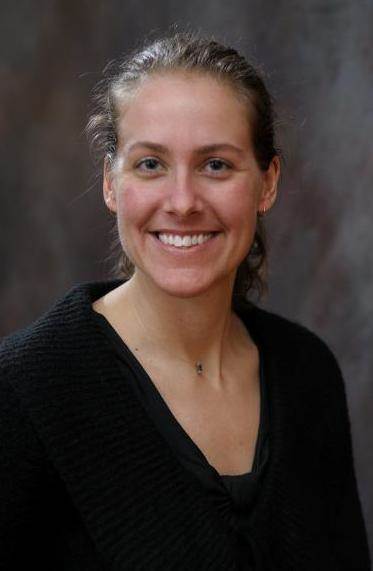}}]{Kira Barton}
(Senior Member, IEEE) received her B.S. degree in Mechanical Engineering from the University of Colorado in 2001, and her M.S. and Ph.D. in Mechanical Engineering from the University of Illinois at Urbana-Champaign in 2006 and 2010. She joined the Mechanical Engineering Department at the University of Michigan, Ann Arbor in 2011. She is currently a Professor in the Robotics Department and Mechanical Engineering Department. She is also serving as the Associate Director for the Automotive Research Center, a University-based U.S. Army Center of Excellence for modeling and simulation of military and civilian ground systems. Prof. Barton’s research specializes in advancements in modeling, sensing, and control for applications in smart manufacturing and robotics. She is the recipient of an NSF CAREER Award in 2014, 2015 SME Outstanding Young Manufacturing Engineer Award, the 2015 University of Illinois, Department of Mechanical Science and Engineering Outstanding Young Alumni Award, the 2016 University of Michigan, Department of Mechanical Engineering Department Achievement Award, and the 2017 ASME Dynamic Systems and Control Young Investigator Award. She was named 1 of 25 leaders transforming manufacturing by SME in 2022, and was selected as one of the 2022 winners of the Manufacturing Leadership Award from the Manufacturing Leadership Council.
\end{IEEEbiography}